\documentclass[10pt]{article}

\usepackage{amsmath}
\usepackage{amssymb}

\usepackage{graphicx}

\usepackage{cite}

\usepackage{color} 

\usepackage[labelfont=bf,labelsep=period,justification=raggedright]{caption}

\makeatletter
\renewcommand{\@biblabel}[1]{\quad#1.}
\makeatother

\date{}

\begin{document}

\begin{flushleft}
{\Large
\textbf{Tactical voting in plurality elections}
}
\\
Nuno A. M. Ara\'ujo$^{1\ast}$, 
Jos\'e S. Andrade, Jr.$^{2}$, 
Hans J. Herrmann$^{1,2}$
\\
\bf{1} Computational Physics for Engineering Materials, Institute for Building Materials, ETH Zurich, Zurich, Switzerland
\\
\bf{2} Departamento de F\'isica, Universidade Federal do Cear\'a, Fortaleza, Cear\'a, Brazil
\\
$\ast$ E-mail: nuno@ethz.ch
\end{flushleft}

\section*{Abstract}

    How often will elections end in landslides and what is the probability for a head-to-head race?
    Analyzing ballot results from several large countries rather anomalous and yet unexplained distributions have been observed. 
    We identify tactical voting as the driving ingredient for the anomalies and introduce a model to study its effect on plurality elections, characterized by the relative strength of the feedback from polls and the pairwise interaction between individuals in the society.
    With this model it becomes possible to explain the polarization of votes between two candidates, understand the small margin of victories frequently observed for different elections, and analyze the polls impact in American, Canadian, and Brazilian ballots.
    Moreover, the model reproduces, quantitatively, the distribution of votes obtained in the Brazilian mayor elections with two, three, and four candidates.

\section*{Introduction}

    The outcome of elections is one of the most stunning phenomena of democracy. 
    Sometimes one candidate wins in a landslide victory and many times two candidates compete head-to-head leaving much suspense. 
    What is the probability of finding a certain situation? 
    This question can be assessed through the distribution of the percentages that each candidate obtains and has in fact been monitored in several large countries with reasonably good statistics. 
    Instead of being Gaussian as one would expect in a simple-minded application of the central limit theorem, an anomalous distribution has been observed. 
    Different mechanisms can affect the electors' opinion formation during an electoral process \cite{Mobilia03,Galam04,Todorov05,Geys06,Castellano09b,Cipra09,Antonakis09}.
    Understanding the effect of such mechanisms poses interesting challenges to both political sciences \cite{Geys06,Brams01,Buchanan04} and statistical physics \cite{Sznajd00,Castellano09}. 
    One distinguishes two different types of elections \cite{Riker82}: {\it proportional} elections, where candidates become elected in proportion to their party's voting fraction \cite{CostaFilho99,CostaFilho03,Bernardes02,Fortunato07}, and {\it plurality} elections, where only the most voted candidate gets a position \cite{Araripe06}. 
    Besides, the latter type can occur on one single ballot or be ``run-off majority'' voting, with two rounds.
    For each type of election different mechanisms were identified \cite{Galam04,Riker82}.
    We focus on plurality elections with one single ballot.
    To describe the tactical strategies applied by electors when $q$ candidates compete against each other, we introduce a model where the competition between pairwise interaction with peers and polls feedback is considered.
          
\begin{figure*}[!ht]
\begin{center}
            \includegraphics[width=\textwidth]{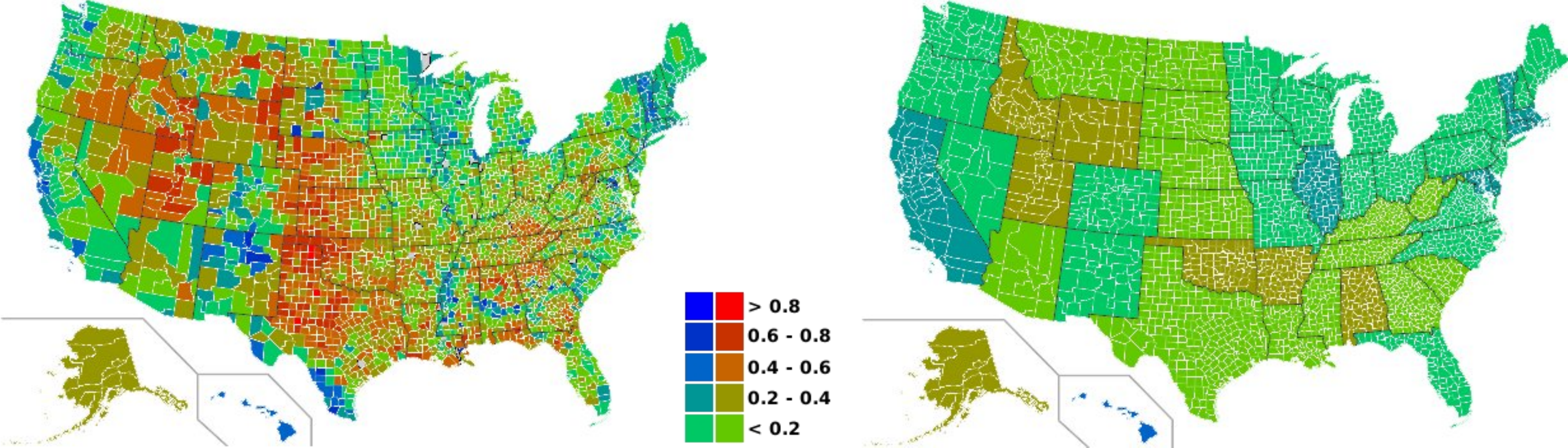} 
\end{center}
        \caption{ 
          Empirical results for the 2008 American presidential elections.
          Maps of the relative difference of votes, $f_v$, in each county (left) and each state (right) showing typical head-to-head runs. 
          Each color represents an interval of $0.2$.
          Blue refers to a landslide victory of Barack Obama and red to a victory of John McCain. 
          Green corresponds to a head-to-head run.
          \label{fig::american.map}
        }
\end{figure*}

\begin{figure}[!ht]
\begin{center}
          \includegraphics[width=0.425\textwidth]{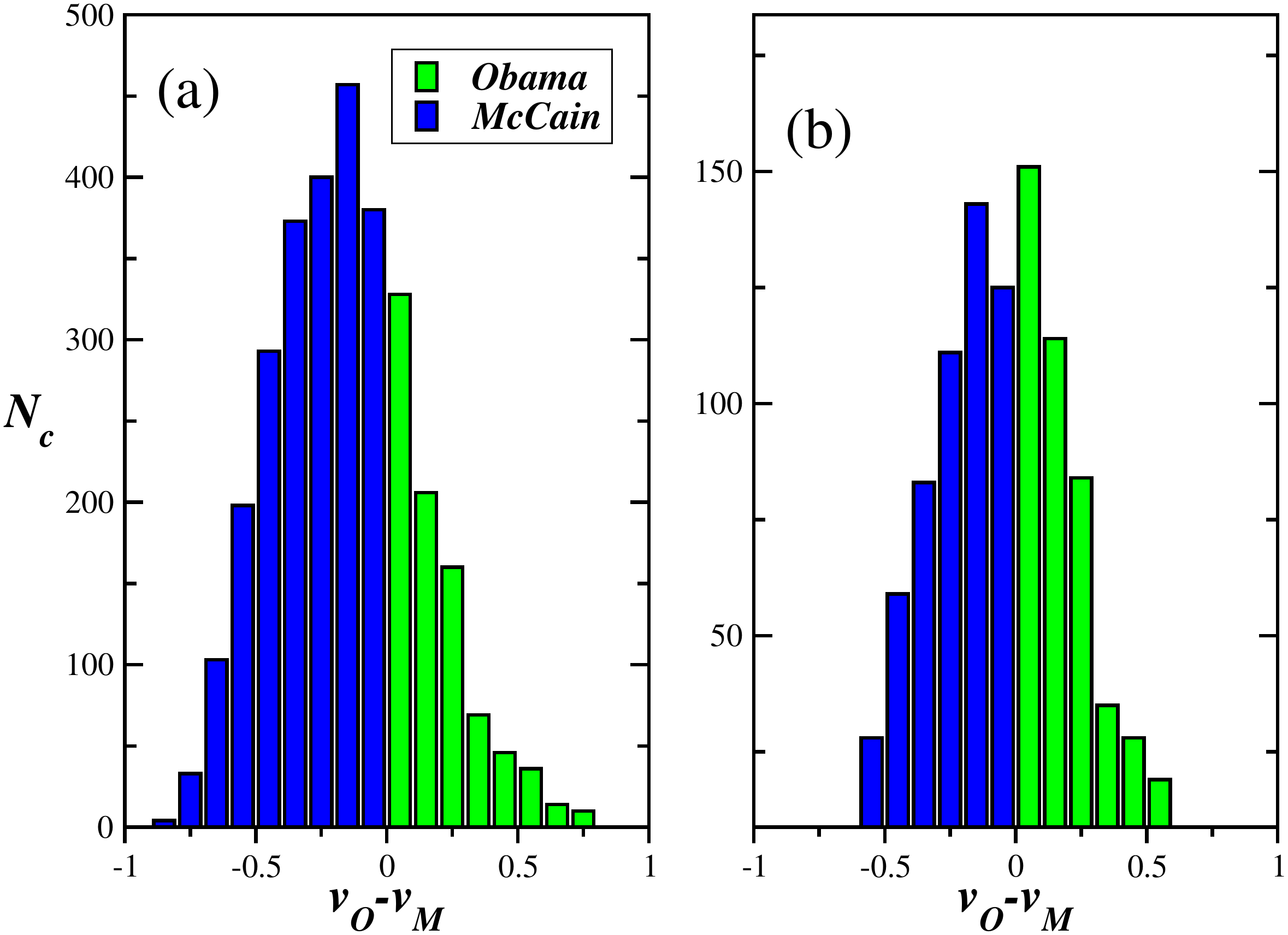}
\end{center}
        \caption{
          Histogram of the relative difference of votes for the 2008 American presidential elections in each county.
          Landslides mostly occur for counties with small number of electors.
          For histogram (a) all counties have been considered and for (b) we have only taken counties with more than $20$ thousand electors.
          \label{fig::american.hist}
        }
\end{figure}

    For plurality elections tactical voting as a response to information from polls is a dominant effect. 
    Specifically, since only the most voted candidate will win, electors can change their vote to an ``unwanted'' candidate just to decrease the margin of victory (also known as ``useful vote''). 
    Maurice Duverger, a French sociologist, recognized this effect as responsible for the emergence of two-party systems in a statement known, today, as Duverger's law \cite{Riker82}.
    Empirical results from different electoral processes also reveal that, typically, victories occur by small margins.
    As an example, let us consider the 2008 American presidential elections.
    The maps of Fig.~\ref{fig::american.map} show the difference in the fraction of votes ($f_v$) between Barack Obama and John McCain per county (left) and per state (right), defined as
    \begin{equation}\label{eq::frac.votes}
      f_v= v_O-v_M \ \ ,
    \end{equation}
    \noindent where $v_O$ and $v_M$ are, respectively, the fraction of votes for the Democratic and Republican candidates.
    For a more quantitative analysis, Fig.~\ref{fig::american.hist} contains the histograms of the differences in the fraction of votes per county, considering the entire set, Fig.~\ref{fig::american.hist}(a), or only for the ones with more than 20 thousand electors, Fig.~\ref{fig::american.hist}(b).
    We observe from these maps and histograms typical head-to-head runs.
    In fact, only in a few counties where the difference in the fraction of votes is slightly above $0.4$, and, at the scale of the state, only margins of victory ($f_v$) below $0.4$ appear.
    From both histograms we conclude that landslides mostly occur for counties with a small number of electors.

    Recently, Restrepo {\it et al.} \cite{Restrepo09} proposed a population dynamics model to study the effect of polls released during the electoral process. 
    They considered two different scenarios: {\it head-to-head} and {\it landslide} voting. 
    In the former, the margin between the first and second candidate is so tight that individuals tend to stick their vote to their favorite candidate. 
    In the latter, the leading candidate has a large margin with respect to all the others, which increases the tendency towards tactical voting.
    To study the effect of the ``useful vote'' on the ballots, we propose a model where a vote results from the competition between two mechanisms: the tendency to align with peers (herding) and the ``useful vote'' \cite{Pattie00}.
    The first one, corresponds to an attractive interaction between somewhat linked electors and solely depends on their opinion.
    The second mechanism, however, depends on the polls and should be proportional to the difference between the fraction of votes of the leading candidate ($v_f$) and the second one ($v_s$). 
    The larger the difference, the stronger the tendency towards tactical voting. 
    Let us consider $q$ candidates identified by an index $\sigma =1,...,q$.
    We propose a balance function $\mathcal{F}$ to quantify the degree of indecision in the society due to the coexistence of different opinions, defined as
      \begin{equation} \label{eq::hamiltonian}
        \mathcal{F}= -H\sum_{<i,j>} \delta(\sigma_i,\sigma_j) + \alpha (v_f-v_s) \sum_i \delta(\sigma_i,f) \ \ ,
      \end{equation}
    \noindent where $\sigma_i$ is the candidate for which citizen $i$ votes and $f$ is the index of the leading candidate (Note that, $v_f$ can only assume values ranging between $1/q$ and unity.). 
    The delta function $\delta(\sigma_i,\sigma_j)$ is unity when $i$ and $j$ chose the same candidate and zero otherwise.
    The herding coefficient $H$ measures the strength of interaction between connected individuals and $\alpha$ the polls impact.
    The natural tendency in society is to minimize its degree of indecision.
    Together with the principle of maximum entropy, this leads to a probability distribution of each state $P\propto\mbox{exp}(-\mathcal{F})$.
    Curiously, in the absence of polls ($\alpha=0$), our model boils down to the so-called $q$-state ``Potts model'' used in magnetism \cite{Wu82}.
    For $H$ above $H_c$ a stable majority opinion can arise, while below it, the interaction between individuals is insufficient to lead to a fixed majority.
    The system is then controlled by the competition between the relative strength of the polls impact, $\alpha/H$, and its degree of subcriticality, $(H-H_c)/H_c$, given by the ratio $R$,
      \begin{equation}\label{eq::defR}
        R=\frac{\alpha/H}{\left(\frac{H-H_c}{H_c}\right)} \ \ .
      \end{equation}
    \noindent The numerator measures the tendency toward tactical voting in the system, the greater $\alpha/H$, the stronger the response to polls.
    This mechanism competes with the emergence of consensus due to pairwise relationships.
    The strength of the psychological coupling between individuals within a society is represented in the model by the social degree of subcriticality.
    Therefore, in our model, for $H$ above $H_c$, in the limit of $R= 0$, when electors decide regardless of the global opinion and only are affected by their peers, rapidly a clear majority appears.
    With increasing $R$, the tendency towards tactical voting becomes more relevant and the margins of victory diminish.
    In the limit of large $R$, the winners' fraction of votes is slightly above $1/q$, i.e., no clear majority evinces.
    For simplicity we consider a square lattice where each node is an elector and solely interacts with its nearest neighbors.
    In fact, as we discuss later, societies are typically better described by small-world networks where long-range connections are also present \cite{Castellano09}.
    
\section*{Results and Discussion}

    To verify the model, let us first consider Brazilian mayor elections since, for such elections detailed data is available \cite{TSEweb} for the number of candidates for each election as well as their percentage of votes.
    Besides, ballot is compulsory and the number of cities and electors large (more than $5500$ cities and $10^8$ electors).
    Since results are resilient over the three elections (2000, 2004, and 2008), we consider, unless otherwise stated, the average over them.

\begin{figure*}
\begin{center}
        \includegraphics[width=0.66\textwidth]{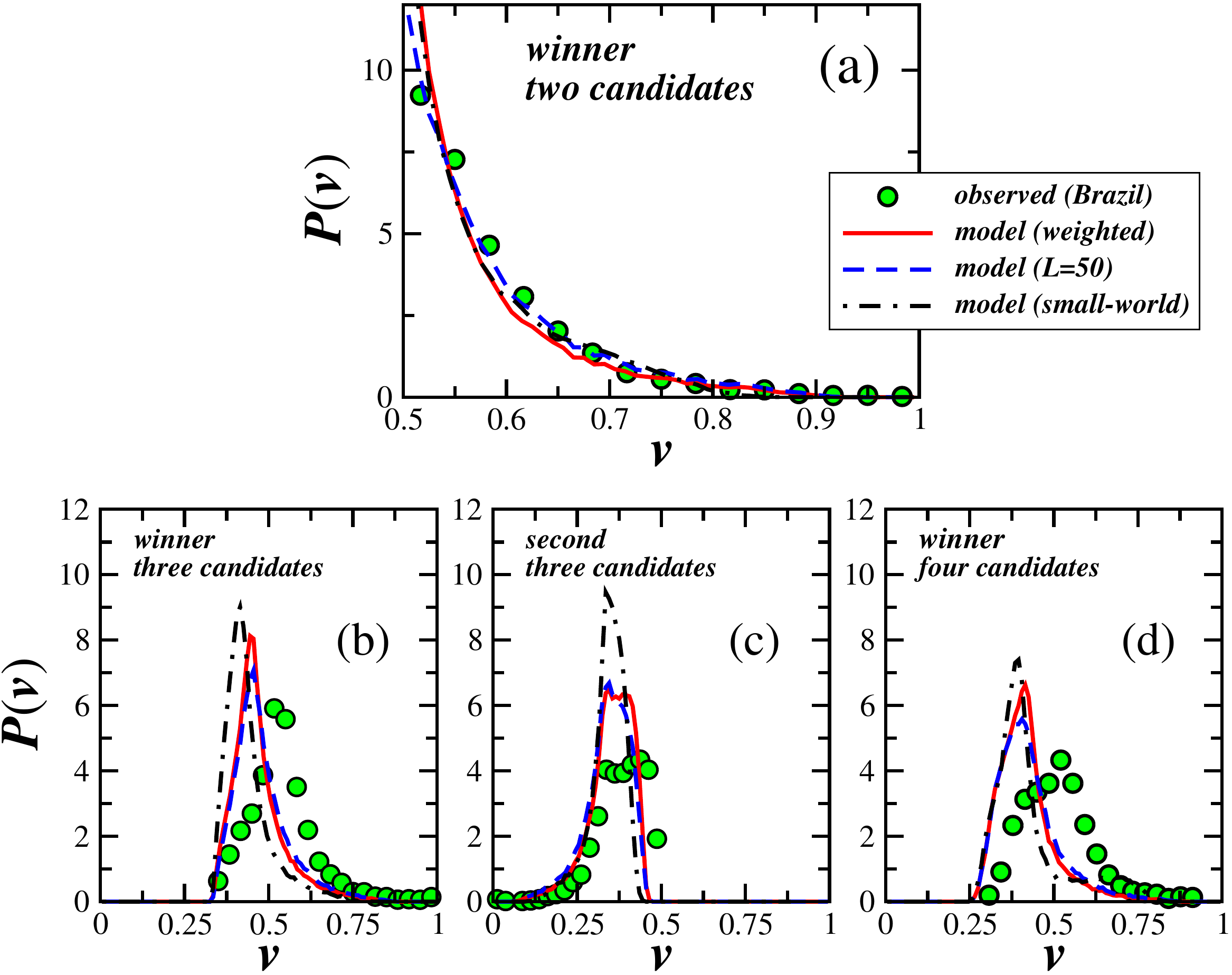}
\end{center}
        \caption{
            Model comparison with empirical results for Brazilian mayor elections.
            Winners' distribution of votes (green) for elections with: a) two, b) three, and d) four candidates.
            Due to the resilience of the 2000, 2004, and 2008 results, we average over these three elections to improve statistics.
            The red-solid lines are the same distributions obtained with the model for $R= 0.69$, by considering a weighted average over three system sizes, namely, $50^2$, $75^2$, and $100^2$ electors.
            The blue-dashed lines correspond to the square lattice with $50^2$ electors.
            The black-dashed-dotted lines are obtained on the small-world topology with $R= 0.46$.
            The second candidate's distributions for the elections with three candidates are included as well (c).
            All results are averages over $5\cdot 10^4$ samples.
          \label{fig::allres}}
\end{figure*}

    Figure~\ref{fig::allres}(a) shows the winners' distribution of votes in the mayor elections with two candidates.
    We use these results to characterize the society in terms of the ratio $R$, of Eq.~(\ref{eq::defR}).
    Fitting the simulational results ($q=2$) to the empirical ones, we obtain $\alpha/H = 0.0069$ and $(H-H_c)/H_c= 0.01$, corresponding to a ratio $R= 0.69$.
    Since the competition involves only two candidates, the winner's percentage of votes is always above $50\%$.
    In fact, most of the winners have a fraction of votes below $70\%$, in agreement with the predicted small margins of victory.
    The real distribution is characterized by an exponential tail \cite{Araripe06} which is also obtained with our model.
    Additionally the model shows that the asymmetry of the distribution increases with $\alpha$.

    Let us consider now the limit of large $R$, i.e., when the strength of the feedback field is much larger than the interaction with peers.
    In this limit the function from Eq.~(\ref{eq::hamiltonian}) can be simplified as,
      \begin{equation}\label{eq::simp.ham}
        \mathcal{F}\approx \alpha Nv_f(v_f-v_s) \ \ ,
      \end{equation}
      \noindent using the equality $\sum_i\delta(\sigma_i,f)= v_f N$, where $N$ is the number of individuals in the system.
    The distribution of the winner's fraction of votes in this limit is then
      \begin{equation}\label{eq::prob.distrib}
        P(v_f)\propto \exp\left[-\alpha N v_f(v_f-v_s)\right] \ \ .
      \end{equation}
    \noindent For the special case of two candidates ($q= 2$), since $v_s=1-v_f$, Eq.~(\ref{eq::prob.distrib}) simplifies as $P(v_f)\propto \exp\left[-\alpha N v_f(2v_f-1)\right]$.
    The resulting dominance of the exponential tail is a trademark of all distributions which has also been pointed out in empirical investigations \cite{Araripe06}.

    Since both the degree of subcriticality and tactical voting impact characterize the society, they should be independent on the number of candidates.
    In Fig.~\ref{fig::allres}(b) we also see the winners' distribution of the fraction of votes for elections with three candidates.
    Results for the candidates ranked as second are included as well, Fig.~\ref{fig::allres}(c).
    Since three candidates are considered, the first candidate has always more than $1/3$ of the total number of votes and the second and third less than $1/2$ and $1/3$, respectively.
    For the winner, a maximum close to $50\%$ is obtained and low margins of victory are observed. 
    As for elections with two candidates, more than $99\%$ of the winners obtain less than $70\%$ of the total votes.
    The red-solid lines represent the simulations for $q= 3$ using the parameters obtained from elections with two candidates.
    A good quantitative agreement between simulational and empirical results is obtained for both candidates (first and second).
    Following Duverger's law, a polarization between two candidates is observed.
    Also in Fig.~\ref{fig::allres}(d), we show results for elections with four candidates.
    Once again, with the same parameters of the two-candidates elections, our model is able to reproduce the empirical results.

    As referred before a square lattice is not very realistic due to the lack of long-range connections.
    In real systems, individuals can contact each other beyond their neighborhood (e.g., by phone, e-mail, virtual-social networks...).
    To account for such a type of connections we also consider a small-world network \cite{Sanchez02}.
    To generate the graph we start with a square lattice and randomly add long-range links until $10\%$ of the individuals have five bonds.
    As seen from the black-dashed-dotted line in Fig.~\ref{fig::allres}, it is again possible to quantitatively reproduce the empirical distributions for all considered numbers of candidates with a single value of $R$, namely, $R= 0.46$.
    Different values of $R$ were obtained for the small-world and regular networks.
    Yet, for each type of topology, when the same value is considered for elections with different number of candidates, the main features of the distribution of votes are recovered.
    The real topologies of interest for opinion dynamics (for example, friendship) are typically small world without scale-free properties \cite{Gonzalez07}.
    For scale-free networks, with an exponent of the degree distribution $\gamma\leq3$, models of opinion dynamics, as well as magnetic ones, are characterized by a clear dominance of an opinion over the other, due to the presence of highly connected nodes \cite{Andrade05,Andrade05b,Andrade09,Castellano09}.
    This is not observed in the empirical results which is consistent with the assumption that topologies underlying opinion interactions are not scale free.

\begin{figure*}
\begin{center}
        \includegraphics[width=0.66\textwidth]{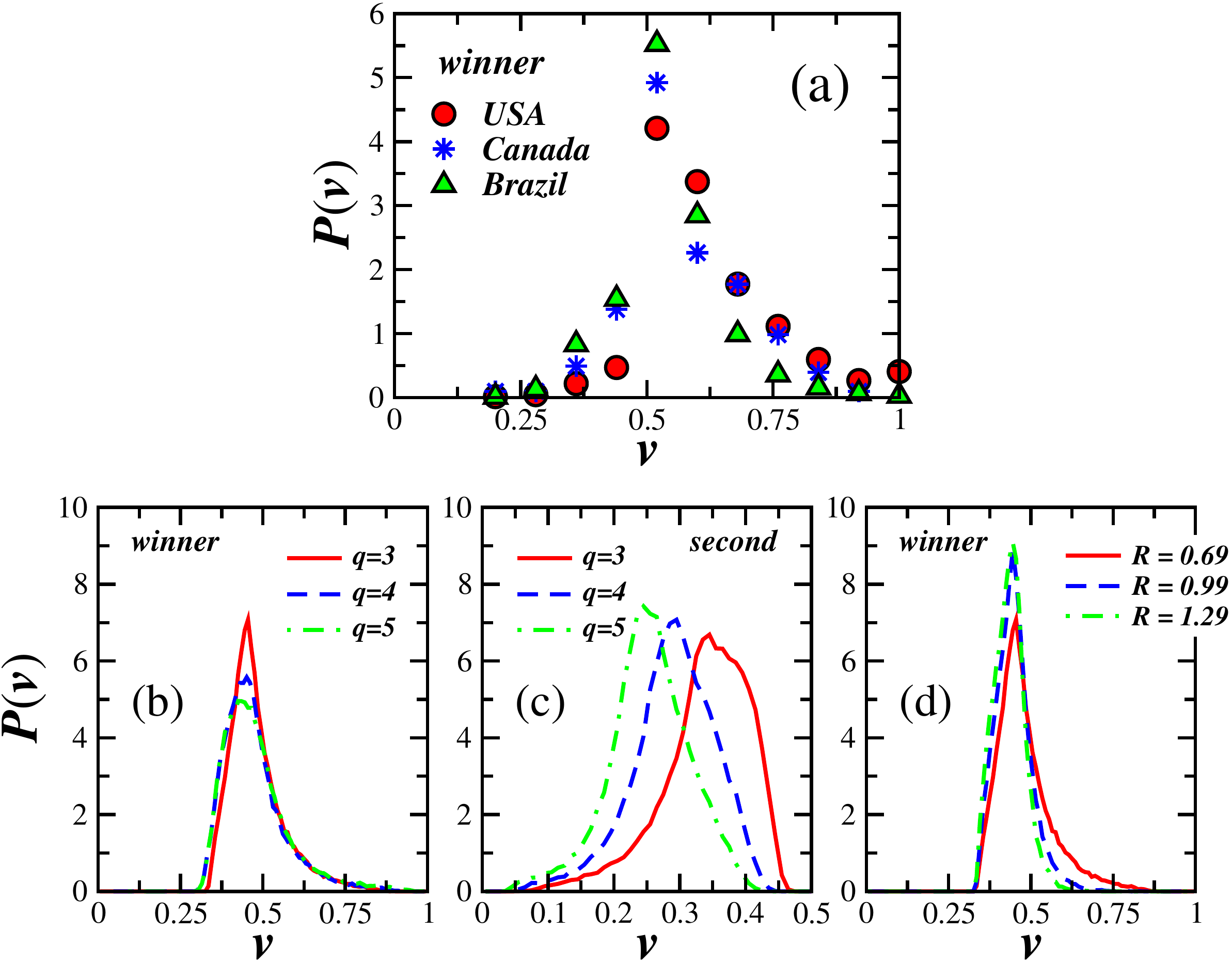}
\end{center}
        \caption{ 
          Comparison between different systems.
          (a) Distribution of the winners' fraction of votes for mayor elections in different countries, namely, United States 2008 (red circles) \cite{USMayorweb}, Canada 2008 (blue stars) \cite{Canadaweb}, and Brazil (green triangles) \cite{TSEweb,Araripe06} where, for the latter one, results have been averaged over three different years 2000, 2004, and 2008, to improve statistics.
          (b) Distribution of the winners' fraction of votes for different number of candidates, namely, three (red-solid line), four (blue-dashed line), and five (green-dotted line). 
          Observed shift with increasing number of candidates has been removed for four and five candidates.
          (c) Distributions for the candidate ranked second.
          (d) Distribution of the winners' fraction of votes for different values of relative strength of the feedback field, for $R= 0.69$ (red-solid line), $0.99$ (blue-dashed line), and $1.29$ (green-dashed-dotted line).
          All simulational results have been averaged over $5\cdot 10^4$ samples on a square lattice with $2500$ electors.
        \label{fig::emp.q.R}}
\end{figure*}

    Let us now consider the results from mayor elections in the United States (2008) \cite{USMayorweb} and Canada (2008) \cite{Canadaweb}.
    In Fig.~\ref{fig::emp.q.R}(a) we show the winners' distribution of votes for these elections together with the ones from Brazil \cite{Araripe06,TSEweb}.
    Regardless of the number of candidates, a maximum is clearly observed for a fraction of votes around $0.5$, with small margins of victory. 
    For most countries the available results are not organized by the number of candidates, therefore, to compare the impact of the feedback field in different countries, it is useful to analyze how the distribution changes with the number of candidates in our model. 
    In Fig.~\ref{fig::emp.q.R}(b) we show this dependence by considering elections with three, four, and five candidates.
    We use always the same value of the relative strength of the feedback field, $R$.
    With increasing $q$ a shift of the peak to lower values occurs, which is also observed in the empirical results for three and four candidates in Ref.~\cite{Araripe06}.
    To compare the tails of the distributions we remove this shift ($0.045$ and $0.09$ for four and five candidates, respectively). 
    We observe that the right-hand side tail of all distributions are the same.
    Besides, a polarization between the first two candidates is observed for all values of $q$ in the inset of Fig.~\ref{fig::emp.q.R}(c).

    To understand how the right-hand side tail is affected by the ratio $R$, we show in Fig.~\ref{fig::emp.q.R}(d) the distributions for different values of $R$ ($0.69$, $0.99$, and $1.29$).
    The larger the strength of the feedback field, $R$, the steeper the right-hand side of the distribution.
    A larger $R$ can be achieved either through a decrease in the degree of subcriticality or an increase in the impact of the feedback field.
    Therefore we can still draw conclusions about the relative strength of the feedback field, through the right-hand side of the distributions for different countries, regardless of the number of candidates.
    From Fig.~\ref{fig::emp.q.R}(a) we observe that the impact of the polls in United States and Canada are very similar.
    Yet, in Brazil the impact reveals to be even stronger.

    Summarizing, to understand plurality elections, we put forward a model where the relevant parameter corresponds to the ratio between the strength of polls impact and society's degree of subcriticality.
    With the model we have been able to compare the distributions from three large countries, namely, United States, Canada, and Brazil.
    The analysis reveals a similar impact of the polls on the elections in United States and Canada and an even stronger one in Brazil.
    Using available results for mayor elections with two candidates in Brazil, we have been able to parametrize the set of electors.
    With the obtained parameters we reproduce the distribution of the fraction of votes in elections with three and four candidates.
    The model has been implemented on a square lattice and a small-world network and, despite the differences discussed previously, the main features of the model do not depend on these topologies, e.g., for both systems, the small margins of victories as well as the polarization between the first two candidates (Duverger's law) are quantitatively reproduced.
    At this stage, our model does not account for blank or null votes, notwithstanding, that such votes shall not affect the candidates' relative fraction of votes.
    Our approach opens up the possibility to make statistical predictions about the outcome of elections by determining the new characteristic parameter $R$ defined as the ratio between polls feedback and herding.
    The next challenge would be to determine this factor $R$ from independent controlled experiments as, for instance, through Internet surveys or questionnaires in schools or neighborhoods.

\section*{Methods}
    We have performed Monte Carlo simulations on the square lattice, for three different system sizes, namely, $50^2$, $75^2$, and $100^2$ electors, and on the small-world network with $50^2$ electors.
    All results have been averaged over $5\cdot10^4$ samples.

\section*{Acknowledgments}
Competence Center for Coping with Crises in Complex Socio-Economic Systems, ETH (Swiss Federal Institute of Technology) Grant 01-08-2, and Swiss National Science Foundation Grant 200021\_126853. Brazilian agencies CNPq (Conselho Nacional de Desenvolvimento Cient\'ifico e Tecnol\'ogico), CAPES (Coordena\c{c}\~ao de Aperfei\c{c}oamento de Pessoal de N\'ivel Superior) and FUNCAP (Funda\c{c}\~ao Cearense de Apoio ao Desenvolvimento Cient\'ifico), and the National Institute of Science and Technology for Complex Systems. The funders had no role in study design, data collection and analysis, decision to publish, or preparation of the manuscript.


\section*{Figure Legends}

\end{document}